\newif\ifspringerclass
\IfFileExists{sn-jnl.cls}{%
  \springerclasstrue
}{%
  \springerclassfalse
}

\ifspringerclass
  \documentclass[pdflatex,sn-basic]{sn-jnl}
\else
  \documentclass[11pt]{article}
  \usepackage[margin=0.9in]{geometry}
\fi
\usepackage[T1]{fontenc}
\usepackage{lmodern}
\usepackage{microtype}
\usepackage{amsmath, amssymb}
\usepackage{graphicx}
\usepackage[labelfont=bf]{caption}
\usepackage{subcaption}
\usepackage{booktabs}
\usepackage{tabularx}
\usepackage{array}
\usepackage[table]{xcolor}
\usepackage{placeins}
\ifspringerclass\else
  \usepackage[authoryear,round]{natbib}
\fi
\ifspringerclass\else
  \usepackage[colorlinks=true, allcolors=blue]{hyperref}
\fi
\usepackage[capitalize,nameinlink,noabbrev]{cleveref}

\definecolor{SWFTableShade}{gray}{0.94}
\graphicspath{{figures/}}

\newcommand{\ObservabilityTitle}{Observability of Finite-Depth Double-Diffusive Exchange from Sparse Temperature-Salinity Measurements}
\newcommand{\ObservabilityShortTitle}{Observability of finite-depth double-diffusive exchange}
\newcommand{\ObservabilityKeywords}{double diffusion; salt fingers; ocean observing systems; hydrographic profiles; Argo; Ice-Tethered Profiler; GO-SHIP}
\newcommand{\ObservabilityAbstractText}{Double-diffusive interfaces can support exchange histories that are spatially organized but only sparsely observed. A resolved three-dimensional calculation contains the route by which scalar gradients broaden, remain compact, connect with remote parts of a finite-depth layer, or organize horizontally, whereas field products usually provide profiles, repeated casts, autonomous-float records, or hydrographic sections. We ask which route-relevant features of finite-depth double-diffusive exchange remain observable after that measurement reduction. Four controlled finite-depth exchange histories are treated as known truth fields and sampled with vertical-profile, profile-bundle, vertical-coarsening, Argo-style, and section-like observing formats. The resulting observables are compared with selected Ice-Tethered Profiler, Argo, and CCHDO/GO-SHIP products to place the synthetic measurements in realistic observing contexts. Isolated profiles are weak route identifiers: at the final comparison time, strict case accuracy remains below 0.48, route-family accuracy is about 0.53--0.56 in leave-one-out profile classification, and 22 of 36 profiles are closer to a different-route profile than to a same-route profile under the current feature set. Profile bundles are substantially more informative. In the bundle-enumeration framework, the one-profile late-time route-family baseline is 0.694, while two-, three-, and four-profile bundles reach route-family accuracies of 0.917, 0.961, and 0.994. Coarse vertical sampling can preserve broad route separation while distorting local interface-width estimates, and section-like sampling adds horizontal-scale information only when station spacing resolves the relevant mode. The useful observing unit for hidden finite-depth double-diffusive exchange is therefore an ensemble or section, not an isolated cast.}

\begin{document}
\ifspringerclass
  \articletype{Research Article}
  \title[\ObservabilityShortTitle]{\ObservabilityTitle}
  \author*[1]{\fnm{Sriram P.} \sur{Kalathoor}}\email{sriram2@gatech.edu}
  \affil*[1]{\orgdiv{Daniel Guggenheim School of Aerospace Engineering}, \orgname{Georgia Institute of Technology}, \orgaddress{\city{Atlanta}, \state{Georgia}, \country{USA}}}
  \abstract{\ObservabilityAbstractText}
  \maketitle
\else
  \title{\ObservabilityTitle}
  \author{Sriram P. Kalathoor\\
\small Daniel Guggenheim School of Aerospace Engineering\\ \small Georgia Institute of Technology\\ \small \texttt{sriram2@gatech.edu}} \date{\today}
  \maketitle
  \begin{abstract}
\ObservabilityAbstractText
  \end{abstract}
\fi

\section{Introduction}

Oceanic temperature-salinity profiles often sample interfaces that are thinner and more structured than the spacing between measurements suggests. In double-diffusive regimes, scalar gradients are not passive markers of stratification alone. They can indicate local susceptibility, organize transport, and preserve part of the history by which an interface has been perturbed and broadened. Classical and modern double-diffusive work has therefore treated temperature-salinity structure as a physically meaningful profile signal as well as a hydrographic descriptor \citep{schmitt1994double}. Profile-based indicators such as Turner angle and water-mass susceptibility maps further show why scalar gradients are routinely read as dynamical clues \citep{ruddick1983practical,you2002turner}. Field evidence also shows that salt-fingering conditions can contribute to diapycnal mixing in oceanic thermoclines \citep{schmitt2005enhanced}. The measurement problem addressed here begins from that profile-visible physics but asks a different question: when an interface has a spatially extended exchange history, how much of that history can a sparse measurement actually see?

The issue is clearest in a finite-depth layer. A local profile can describe the scalar transition at the instrument location, but the same profile does not reveal whether the transition came from a compact interface, a vertically extended connection, a short-scale horizontally organized route, or a mixed route whose details depend on phase realization. These route histories differ in horizontal organization and vertical reach, so their signatures are not fully contained in one temperature-salinity cast. A field observer may see a sharp gradient, a broadened interface, or a locally correlated temperature and salinity anomaly, while the hidden field contains a broader pattern of exchange.

This mismatch between full-field route history and sparse measurement is the central premise of the analysis. The finite-depth simulations used here are treated as known truth fields. The route-selection physics is taken from the companion finite-depth study \citep{kalathoor2026roughnessPreprint}; the present question is what a profile, a small profile bundle, a coarsened profile, an autonomous-profile sampling pattern, or a hydrographic section would recover from those known histories. The simulation truth gives access to the hidden route label; the synthetic observation gives the measurement that would be available to an observer.

Real ocean products enter as observing-format context. Ice-Tethered Profilers provide high-vertical-resolution profiles under Arctic pack ice \citep{krishfield2008automated,toole2011itp}, and ITP records have been used to characterize Arctic thermohaline staircase structure \citep{timmermans2008ice,shibley2017arctic}. Argo provides routine autonomous profile sampling and quality-control structure \citep{roemmich2009argo,roemmich2019future}, including profile archives that have been used to build global thermohaline-staircase datasets \citep{vanderboog2021global}. CCHDO/GO-SHIP products provide section-scale hydrographic sampling with station spacing and vertically dense CTD casts \citep{sloyan2019goship}. These records are not used as direct validation targets for the simulated salt-fingering route histories. The Arctic staircase branch, for example, provides a branch-aware contrast instead of a salt-fingering route match \citep{kelley2003diffusive,radko2013double}. Their role here is to anchor the measurement reductions in formats that field programs actually use.

The analysis makes four linked contributions. First, it quantifies the ambiguity of isolated temperature-salinity profiles when the hidden finite-depth route is known. Second, it shows how small bundles of profiles recover route-family information that a single cast loses. Third, it separates broad route-state observability from quantitative interface-metric fidelity under vertical coarsening. Fourth, it places the same scalar-profile and section quantities beside selected ITP, Argo, and CCHDO/GO-SHIP products, yielding a measurement-design hierarchy built from route-known simulation truth.

\section{Truth Set, Observing Reductions, and Observable Quantities}

\subsection{Finite-Depth Truth Fields}

The truth set consists of four finite-depth simulations with identical background control parameters and different initial horizontal spectral content or random realization. The histories are identified by their observable route character. The high-annulus route begins from shorter horizontal scales and retains a compact, fine-scale scalar interface for much of the evolution. The low-mode route emphasizes broad horizontal organization and develops a more vertically extended scalar transition. The mixed baseline combines low- and high-wavenumber content. The mixed seed realization shares the same broad mixed route family but differs in detailed phase realization.

The observability calculation uses saved scalar slices from these completed runs. The fields are fixed inputs: a controlled set of known exchange histories. Every sampled profile or section inherits a hidden route label from the truth field, but the observable metrics are computed without using that label. This separation allows the analysis to ask whether a sampled measurement contains enough information to recover a route family, while keeping the route family itself as an external truth label.

\subsection{Synthetic Profile Sampling}

\begin{figure}[!htbp]
  \centering
  \includegraphics[width=\linewidth]{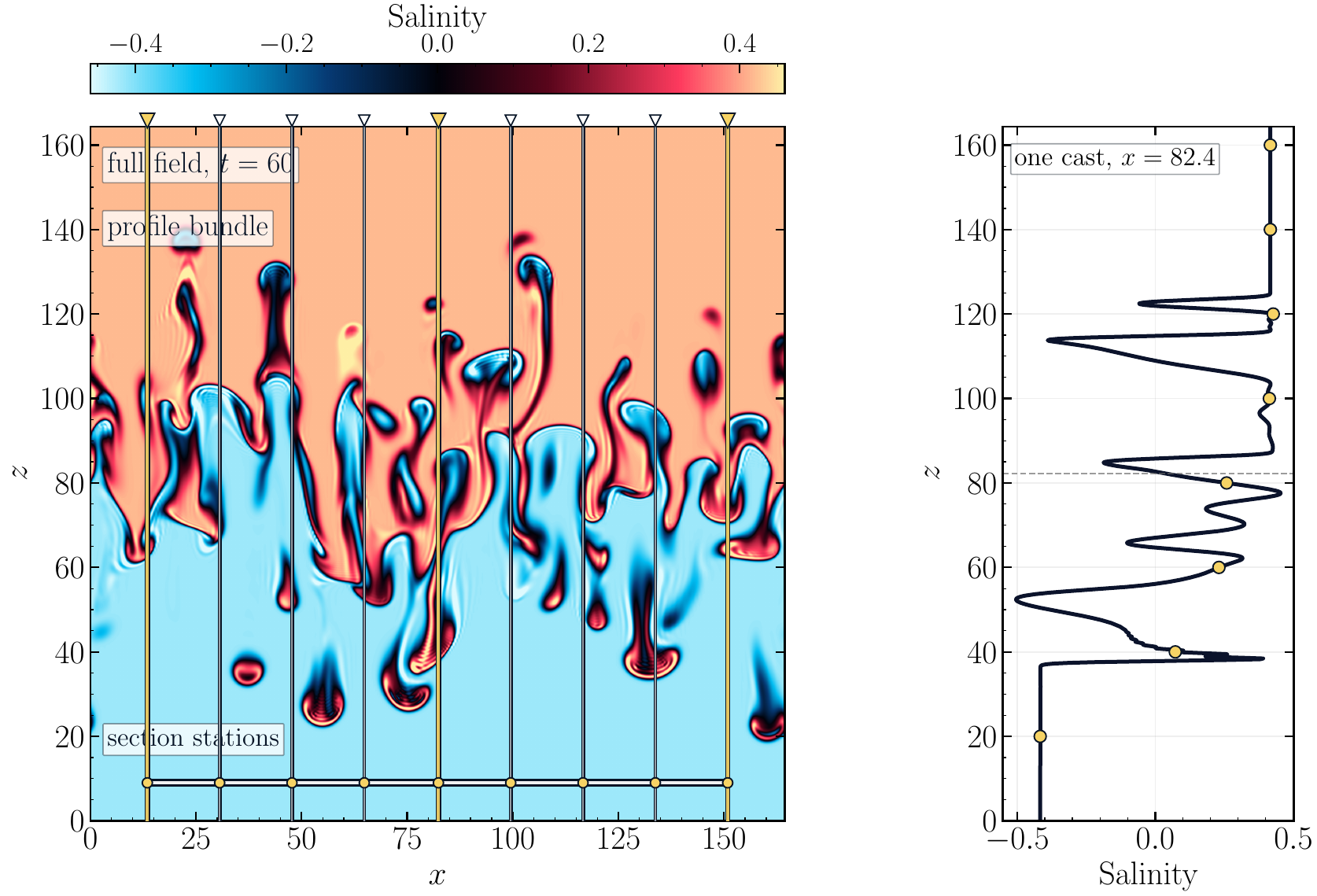}
  \caption{Measurement reduction from a finite-depth scalar field to observable profile and section products. The full field contains the hidden exchange route, while profiles and section stations retain only restricted parts of that route history.}
  \label{fig:bridge}
\end{figure}

The first reduction is a set of vertical temperature-salinity profiles sampled from the simulated field at selected horizontal locations and output times. For each route history, nine profile locations are sampled from the mid-plane slice. Each profile supplies scalar-gradient, scalar-correlation, and interface-location information. \Cref{fig:bridge} introduces this reduction by showing a field, profile locations, section-like station locations, and one derived profile. The physical object is a field; the observation is a sparse projection of that field.

Profile bundles are formed by grouping multiple profiles from the same route history and comparison time. A single profile asks whether one local cast can identify a route. A bundle asks whether a small spatial ensemble contains enough information to recover route-family structure or stabilize scalar-interface metrics. This difference between a local cast and an ensemble is the main measurement-design lever in the profile analysis.

\subsection{Observable Metrics}

The profile observables are deliberately narrower than the full simulation state. They are quantities that can be computed from sampled temperature-salinity profiles: salinity-gradient interquartile width, central salinity-gradient fraction, outer salinity-gradient fraction, gradient entropy, and profile temperature-salinity correlation. These quantities describe how a sampled scalar transition is distributed in the vertical and how tightly temperature and salinity co-vary within the sampled interval.

The salinity-gradient interquartile width measures the vertical span occupied by the central half of the salinity-gradient weight. A compact interface has a smaller width, while a broadened or split interface has a larger width. The central fraction measures how much salinity-gradient weight remains near the nominal interface. The outer fraction measures how much has migrated away from that central region. Gradient entropy measures how spread out the gradient weight is across the sampled window. The profile temperature-salinity correlation measures local scalar coupling.

These profile quantities are useful but limited. They can describe the sampled interface, but they do not by themselves measure flux, identify the full exchange route, or recover horizontal organization. Section observables are introduced when horizontal position is available. The section dominant mode measures the leading horizontal scale of the sampled interface track, and section roughness measures the amplitude of interface-position variation across stations. \Cref{tab:profile-observable-metrics,tab:route-section-observable-metrics} define the observables, required inputs, physical meanings, and claim boundaries.

\begin{table}[!htbp]
  \centering
  \caption{Profile quantities used to reduce resolved finite-depth fields to local scalar measurements.}
  \label{tab:profile-observable-metrics}
  \begingroup
\small
\setlength{\tabcolsep}{3pt}
\renewcommand{\arraystretch}{1.08}
\begin{tabularx}{\linewidth}{
  >{\raggedright\arraybackslash}p{0.17\linewidth}
  >{\raggedright\arraybackslash}p{0.19\linewidth}
  >{\raggedright\arraybackslash}p{0.22\linewidth}
  >{\raggedright\arraybackslash}p{0.19\linewidth}
  >{\raggedright\arraybackslash}X}
\toprule
Metric & Required input & Physical meaning & Supports & Does not support \\
\midrule
\rowcolor{SWFTableShade}
Interface center \(z_c\) &
Profile or section \(T,S,z\) &
Gradient-weighted center of the sampled scalar interface &
Interface location and metric alignment &
Hidden-route identification from one cast \\
Salinity-gradient IQR width \(W_S\) &
Profile \(S(z)\) &
Vertical width containing the central half of salinity-gradient weight &
Compact versus broadened scalar-interface state &
Full exchange flux or three-dimensional plume geometry \\
\rowcolor{SWFTableShade}
Central gradient fraction \(C_S\) &
Profile \(S(z)\) and an interface-centered inner band &
Fraction of gradient weight retained near the interface &
Near-interface scalar-gradient concentration &
Finite-depth connection by itself \\
Outer gradient fraction \(O_S\) &
Profile \(S(z)\) and an interface-centered outer band &
Fraction of gradient weight displaced away from the interface core &
Outer-layer scalar-gradient spread &
Exchange direction or pathway without ensemble context \\
\rowcolor{SWFTableShade}
Gradient entropy \(H_S\) &
Profile \(S(z)\) &
Spread of normalized salinity-gradient weight over the profile &
Scalar-interface complexity and distributed gradients &
Transport strength or route identity alone \\
Profile correlation \(r_{TS}\) &
Co-sampled \(T(z)\) and \(S(z)\) &
Linear scalar co-variation in the sampled interval &
Scalar coupling in the observed profile &
Branch or route validation without regime context \\
\bottomrule
\end{tabularx}
\endgroup

\end{table}

\begin{table}[!htbp]
  \centering
  \caption{Route-family and section quantities used to assess ensemble and horizontal observability.}
  \label{tab:route-section-observable-metrics}
  \begingroup
\small
\setlength{\tabcolsep}{3pt}
\renewcommand{\arraystretch}{1.08}
\begin{tabularx}{\linewidth}{
  >{\raggedright\arraybackslash}p{0.17\linewidth}
  >{\raggedright\arraybackslash}p{0.20\linewidth}
  >{\raggedright\arraybackslash}p{0.22\linewidth}
  >{\raggedright\arraybackslash}p{0.19\linewidth}
  >{\raggedright\arraybackslash}X}
\toprule
Metric & Required input & Physical meaning & Supports & Does not support \\
\midrule
\rowcolor{SWFTableShade}
Route-family classification &
Standardized profile or bundle metrics from the controlled truth set &
Assignment to broad route family &
Synthetic route-family observability as bundle size changes &
Universal classification of arbitrary field profiles \\
All-metric stable fraction &
Profile-bundle metrics compared with full-bundle values &
Fraction of bundles reproducing all selected scalar metrics &
Profile count needed for quantitative metric stability &
Stable plume timing or full three-dimensional transport \\
\rowcolor{SWFTableShade}
Section dominant mode \(m_*\) &
Interface-center track along a section &
Leading resolved horizontal mode of interface displacement &
Horizontal scale recovery when station spacing resolves the mode &
Modes above the section Nyquist limit \\
Section roughness \(R_z\) &
Interface-center track along a section &
RMS or standard deviation of interface displacement &
Section-sampled interface displacement amplitude &
Complete spectral-memory state without mode or band information \\
\bottomrule
\end{tabularx}
\endgroup

\end{table}

\subsection{Real Observing Formats}

The real observing products enter at three levels. The ITP product provides a high-resolution profile context. The retained ITP set contains 166 upper-interface profiles after screening and supplies a real distribution of interface widths, entropy, and temperature-salinity correlations. The Argo pass supplies autonomous-profile sampling records with quality-control flags, pressure spacing, and interface-centered sampling. The screened Argo set contains 54 candidate profile records and yields 36 accepted local sampling records drawn from 22 unique profiles. The CCHDO/GO-SHIP I06S product supplies section context with 109 profiles, 108 usable candidate-interface profiles, a median station spacing of 55.4 km, and a median strict-QC pressure spacing of 2.0 dbar.

\Cref{tab:observing-formats} records how these products are used. ITP provides dense real-profile context. Argo provides routine autonomous-profile sampling realism. CCHDO/GO-SHIP provides station-spacing and vertical-sampling context for section observability. The distinction is essential: the real products anchor observing formats, while the finite-depth simulations provide the route-known truth fields.

\begin{table}[!htbp]
  \centering
  \caption{Observing-format roles used to connect the synthetic reductions to real profile and section products.}
  \label{tab:observing-formats}
  \small
  \setlength{\tabcolsep}{2pt}
  \renewcommand{\arraystretch}{1.12}
  \begin{tabular}{p{0.19\linewidth}p{0.24\linewidth}p{0.28\linewidth}p{0.22\linewidth}}
\toprule
Observing format & Sampling scale used here & Use in analysis & Observational limitation \\
\midrule
ITP 84 upper-interface profiles & 166 retained profiles; median native spacing 1.0 dbar; local window +/-82.1 dbar & High vertical-resolution profile context for interface-width, entropy, and T-S correlation metrics. & Single-column profiles do not give simultaneous horizontal connectivity or route identity. \\
Argo-style autonomous profiles & 36 accepted local sampling records drawn from 22 unique profiles; median 82.5 levels; median spacing 1.98 dbar & Tests how autonomous profile sampling preserves or degrades route-relevant profile observables. & Profiles are sparse in space and time; the operator cannot see planform organization directly. \\
CCHDO/GO-SHIP I06S section & 109 stations over 4540 km; median station spacing 55.4 km; median vertical spacing 2.0 dbar & Field-scale section analogue for horizontal sampling, interface tracking, and station-spacing limits. & Station spacing is much coarser than short salt-finger planform scales, and one section does not resolve time evolution. \\
CCHDO/GO-SHIP interface track & 108 usable interface profiles; median interface pressure 88.6 dbar; median width 45.3 dbar & Demonstrates that the same gradient-based interface bookkeeping can be applied to a real hydrographic section. & Useful for section-scale context, not for validating the hidden route taken by the idealized simulation. \\
\bottomrule
\end{tabular}

\end{table}
\FloatBarrier

\section{One Profile, Many Profiles, and the Recovery of Route Family}

\subsection{Isolated Profiles Collapse Distinct Routes}

The first result is intentionally negative. The single-profile test asks whether one temperature-salinity cast can identify the hidden route from profile observables alone. Two leave-one-out classifiers were applied in standardized observable space: a nearest-centroid classifier and a three-nearest-neighbor classifier. The target was evaluated as both strict case identity and a broader route family in which the two mixed realizations are grouped together.

At the final comparison time, the strict case accuracy remains below 0.48. Route-family accuracy is only 0.528 for nearest centroid and 0.556 for three-nearest-neighbor classification. The nearest-profile overlap test gives the same conclusion geometrically: 22 of 36 profiles are closer to a profile from a different route than to one from the same route under the current feature set. \Cref{fig:profile-observability-single} shows this single-profile ambiguity.

Profiles remain informative local measurements. A profile measures local scalar-gradient width, central concentration, outer-gradient fraction, entropy, and scalar correlation. The limitation is that these local quantities do not uniquely encode the spatial route by which the finite-depth field arrived at that profile state. A compact route and a mixed route can produce overlapping local profile features; a broadened local interface can be the visible cut through more than one hidden exchange history.

\subsection{Profile Bundles Convert a Local Problem Into an Ensemble Problem}

The bundle calculation changes the observing unit from a single cast to a small ensemble. All possible bundles were enumerated from the nine profile locations at each selected time. The same standardized feature space was used, but each bundle samples more of the horizontal variation that distinguishes route families.

\begin{figure}[!htbp]
  \centering
  \begin{subfigure}[t]{0.49\linewidth}
    \centering
    \includegraphics[width=\linewidth]{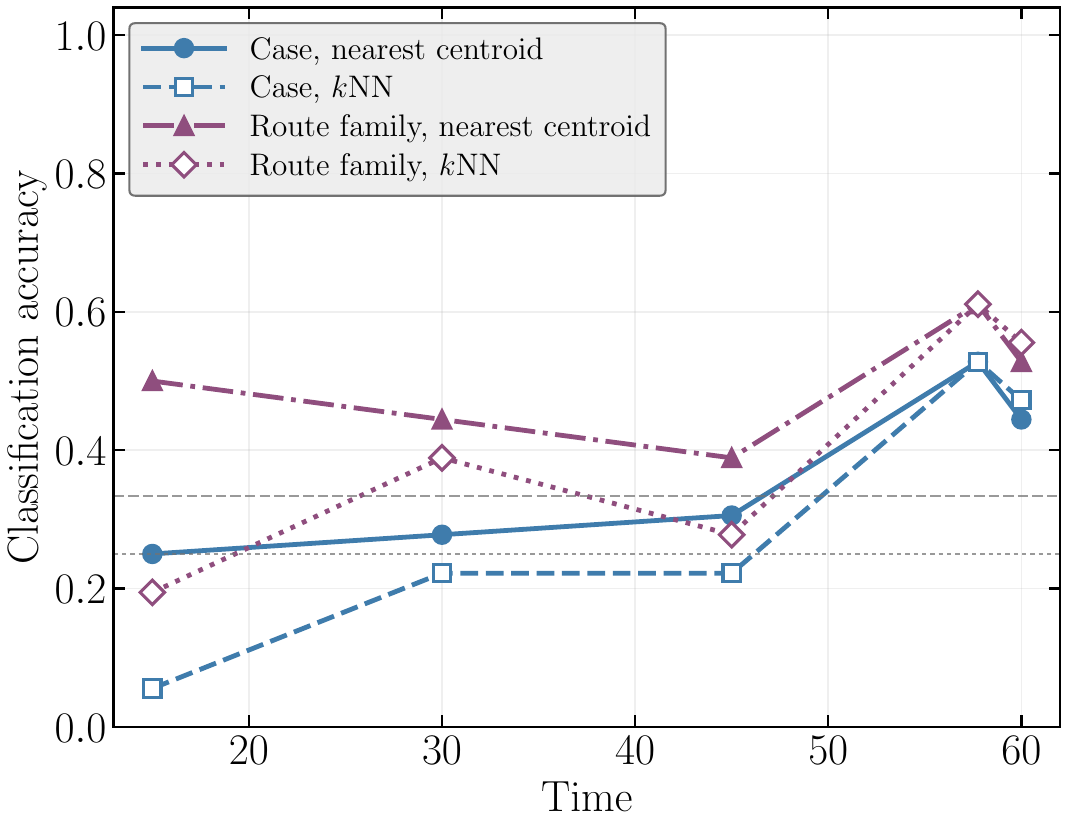}
    \caption{Single-profile route recovery.}
    \label{fig:profile-observability-single}
  \end{subfigure}
  \hfill
  \begin{subfigure}[t]{0.49\linewidth}
    \centering
    \includegraphics[width=\linewidth]{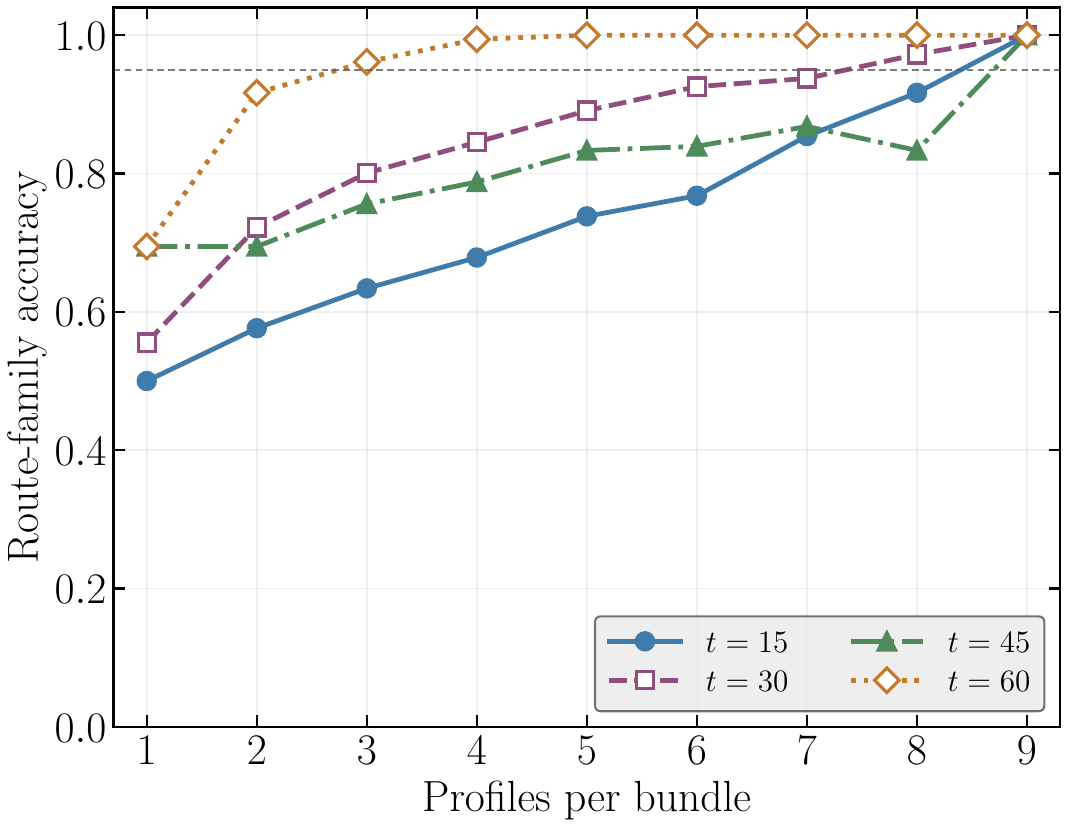}
    \caption{Profile-bundle route-family recovery.}
    \label{fig:profile-observability-bundle}
  \end{subfigure}
  \caption{Profile observability of hidden route family. Isolated profiles remain ambiguous, while small profile bundles recover route-family structure much more reliably in the controlled truth set.}
  \label{fig:profile-observability}
\end{figure}

The improvement is large by the final comparison time. In the bundle-enumeration framework, the one-profile route-family baseline is 0.694. That one-profile baseline is higher than the independent leave-one-out classifier values because the tests are different, but it still leaves nearly one third of one-profile bundle samples misclassified. Adding a second profile raises route-family accuracy to 0.917. Three profiles reach 0.961, four profiles reach 0.994, and five profiles reach 1.000 in the current controlled truth set. \Cref{fig:profile-observability-bundle} shows the bundle-size recovery.

The bundle result is a constructive measurement-design result. The two- or three-profile threshold is specific to this controlled truth set, but the mechanism is general to spatially structured measurements: aggregation changes the observable problem. A single profile samples one local scalar transition. A bundle samples part of the spatial organization that separates compact, broad, and mixed routes.

\subsection{Route-Family Recovery Precedes Quantitative Metric Stability}

Route-family classification and quantitative metric stability improve at different rates. The all-metric stable fraction remains 0.000 for a single profile at the final comparison time. It rises only to 0.089 for three profiles, reaches 0.889 for eight profiles, and reaches 1.000 for nine profiles. Thus a small bundle can identify broad route-family differences before it can reproduce all scalar-interface metrics to the strict tolerances used here.

This separation matters for observation planning. If the goal is to distinguish broad route families, a small local ensemble may be sufficient in a favorable setting. If the goal is to estimate interface width, outer-gradient fraction, entropy, and scalar correlation quantitatively, more profiles are needed. Route-family identification and local interface-state fidelity are therefore related but distinct measurement goals. \Cref{tab:profile-observability} summarizes the time-dependent single-profile and profile-bundle thresholds that support this distinction.

\begin{table}[!htbp]
  \centering
  \caption{Profile-observability summary across comparison times. Single-profile values report route-family accuracy from the nearest-centroid classifier. Bundle-size values give the minimum number of profiles required to reach at least 0.95 route-family accuracy in the bundle-enumeration framework.}
  \label{tab:profile-observability}
  \begin{tabular}{lrrrr}
\toprule
Quantity & $t=15$ & $t=30$ & $t=45$ & $t=60$ \\
\midrule
Single-profile route-family accuracy & 0.50 & 0.44 & 0.39 & 0.53 \\
Profiles for $\ge 0.95$ route-family accuracy & 9 & 8 & 9 & 3 \\
\bottomrule
\end{tabular}

\end{table}

\FloatBarrier

\section{Vertical Resolution and the Fidelity of Profile Observables}

\subsection{Coarsening Degrades Interface-Width Estimates}

Vertical coarsening was applied to synthetic and real profile products to determine how much metric fidelity is lost when the vertical coordinate is sampled more sparsely. The clearest effect is on salinity-gradient interquartile width. At 20-unit profile-coordinate spacing, the median relative error in salinity-gradient IQR is 0.436 for the retained ITP profile set, 0.372 for the high-annulus route, 0.239 for the low-mode route, 0.407 for the mixed baseline, and 0.340 for the mixed seed realization. \Cref{fig:resolution-degradation-iqr} shows these errors.

These errors are large enough that coarse profiles are poor quantitative interface-width measurements. The error changes how much scalar-gradient weight appears to be concentrated near the interface and how much appears to occupy the outer parts of the profile. In an interface-controlled exchange problem, those quantities determine whether the sampled transition appears compact, broadened, or connected to remote parts of the layer.

\subsection{Broad State Separation Can Persist After Coarsening}

\begin{figure}[!htbp]
  \centering
  \begin{subfigure}[t]{0.49\linewidth}
    \centering
    \includegraphics[width=\linewidth]{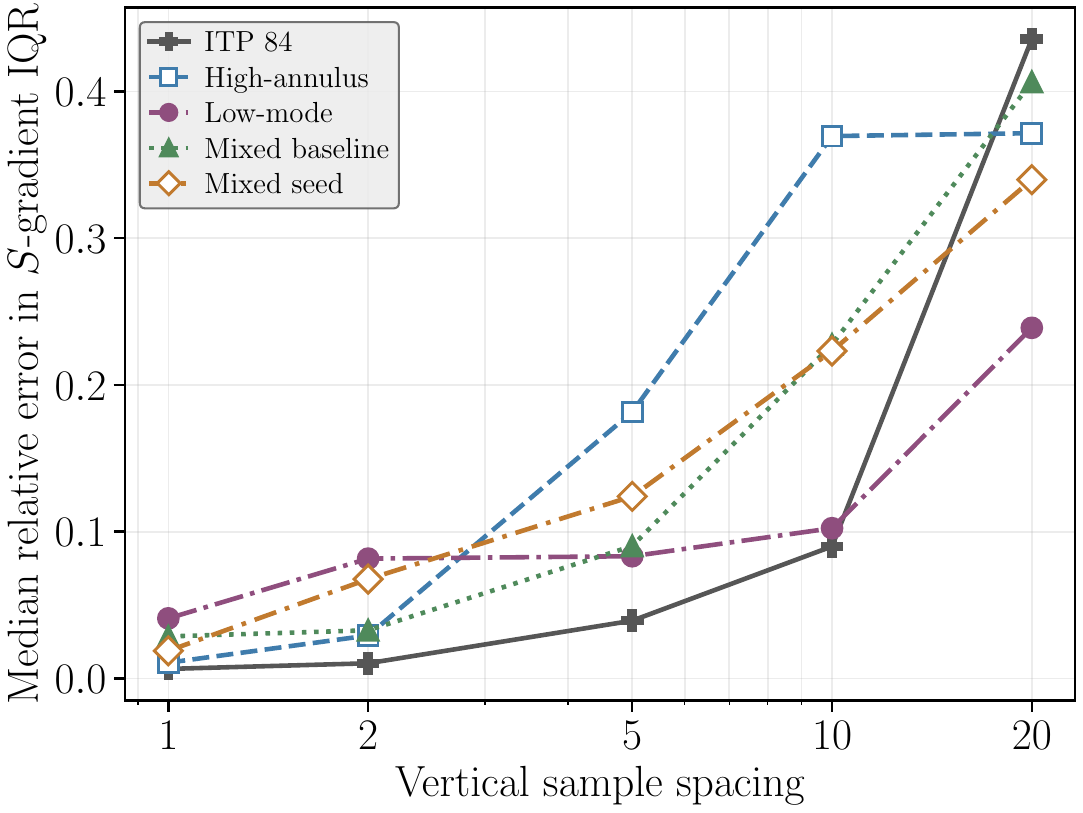}
    \caption{Interface-width error.}
    \label{fig:resolution-degradation-iqr}
  \end{subfigure}
  \hfill
  \begin{subfigure}[t]{0.49\linewidth}
    \centering
    \includegraphics[width=\linewidth]{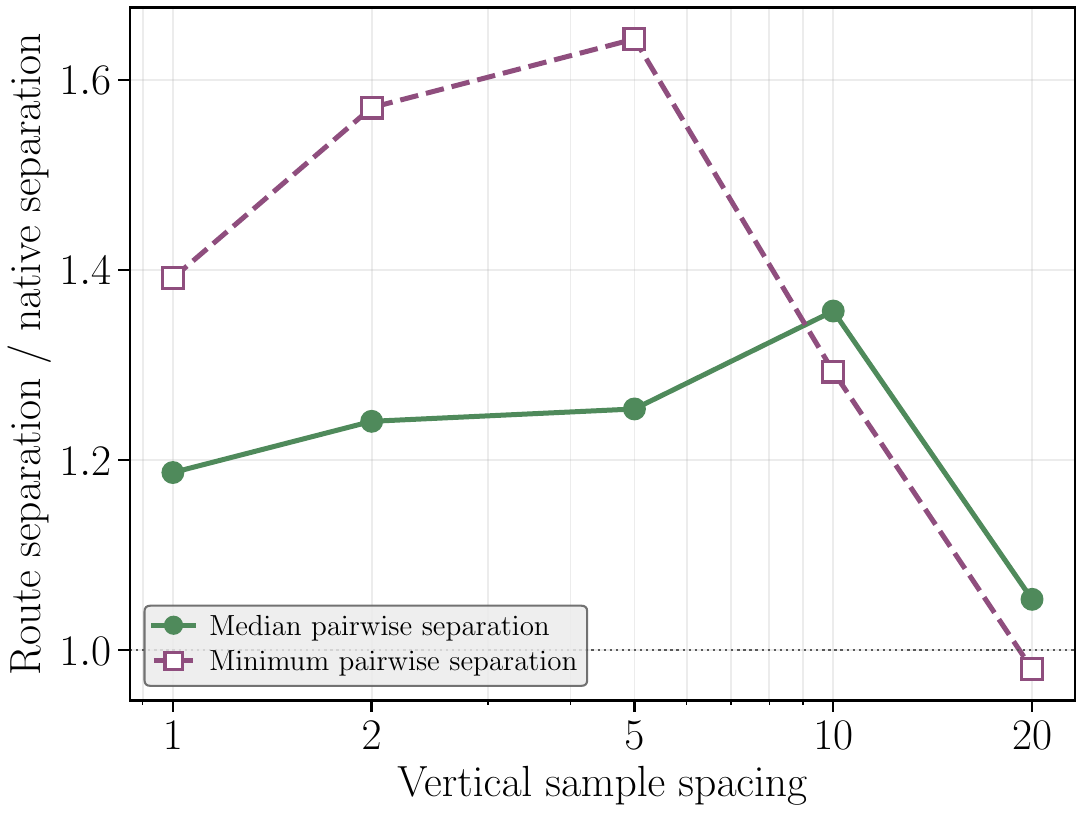}
    \caption{Retained route separation.}
    \label{fig:resolution-degradation-separation}
  \end{subfigure}
  \caption{Vertical-resolution effects on observable quantities. Coarsening can distort local interface-width estimates while preserving broad route-state separation in the reduced feature space.}
  \label{fig:resolution-degradation}
\end{figure}

The coarsening result also has a constructive side. Although local interface metrics are distorted, broad separation among route states can remain visible in the standardized observable space. At the final comparison time, median pairwise route separation is 1.357 of the native value at 10-unit spacing and 1.053 of the native value at 20-unit spacing; the minimum pairwise separation at 20-unit spacing is 0.980 of native. \Cref{fig:resolution-degradation-separation} shows this retained route separation.

A retained separation ratio near or above one indicates that the particular reduced feature set preserves broad state differences. The same coarse profiles can still distort the physical width or outer-gradient estimate that an interface interpretation would need. Broad state observability therefore survives more easily than quantitative metric fidelity.

\FloatBarrier

\section{Real High-Resolution Profiles as Branch-Aware Context}

The real-profile comparison asks whether the same scalar bookkeeping can be applied to ocean profiles, and how those real profiles sit relative to the synthetic route states. The retained ITP set contains 166 upper-interface profiles. The median interface-center pressure is 37 dbar. The median salinity-gradient IQR is 23.499, the median central fraction is 0.570, the median outer fraction is 0.185, the median salinity-gradient entropy is 0.857, and the median temperature-salinity correlation is 0.619.

\begin{figure}[!htbp]
  \centering
  \begin{subfigure}[t]{0.49\linewidth}
    \centering
    \includegraphics[width=\linewidth]{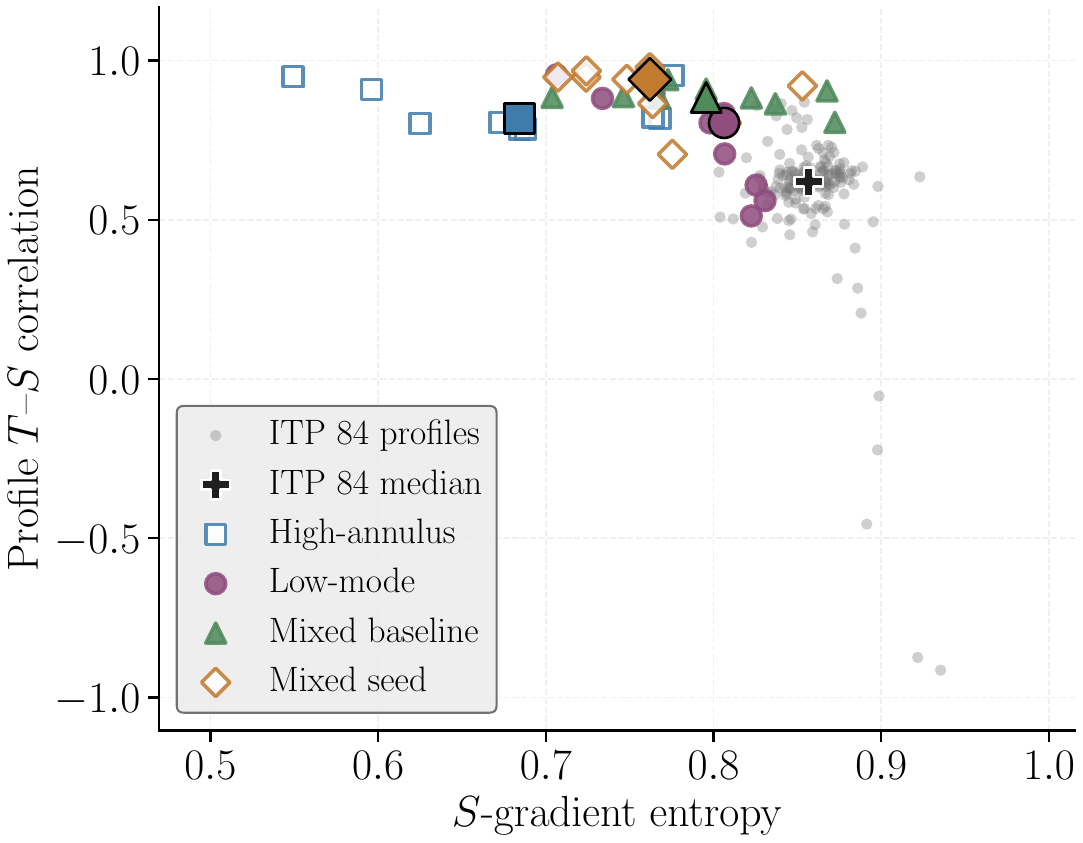}
    \caption{Entropy versus scalar correlation.}
    \label{fig:observable-state-entropy}
  \end{subfigure}
  \hfill
  \begin{subfigure}[t]{0.49\linewidth}
    \centering
    \includegraphics[width=\linewidth]{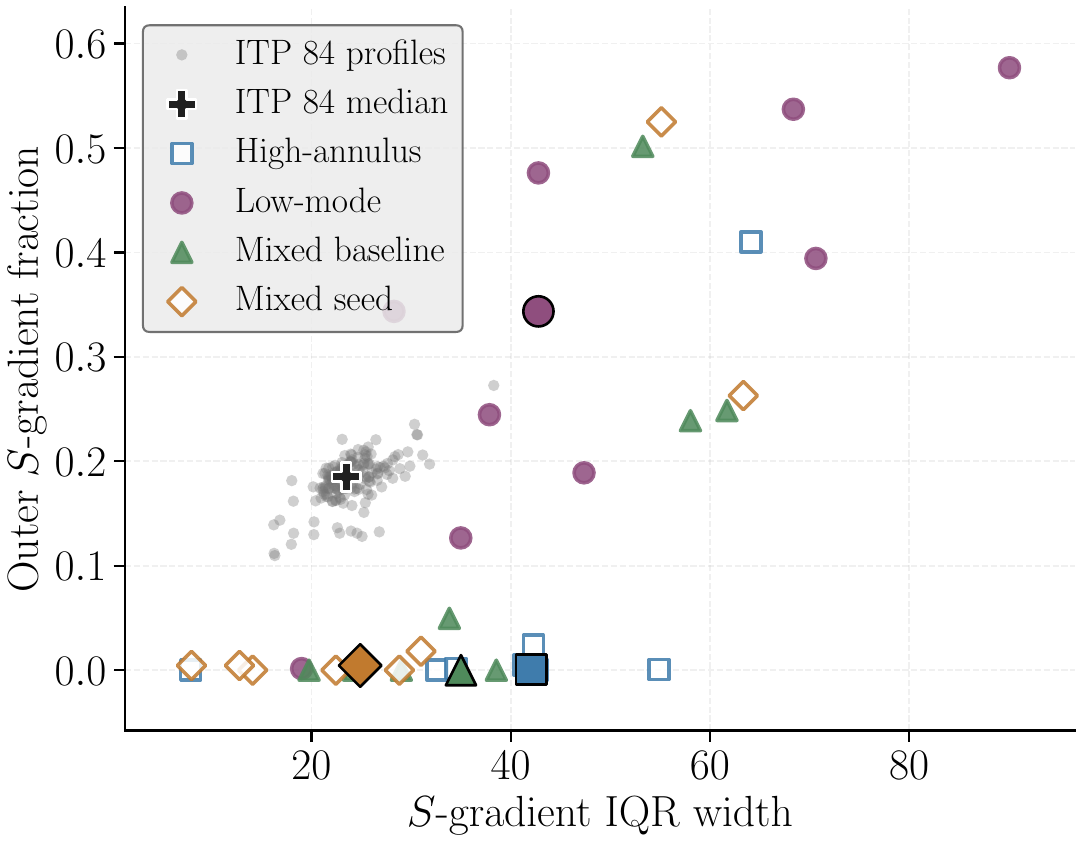}
    \caption{Interface width versus outer gradient fraction.}
    \label{fig:observable-state-width}
  \end{subfigure}
  \caption{Real high-resolution profile context in the same observable coordinates as the synthetic profile reductions. The comparison is branch-aware: it tests shared scalar bookkeeping without requiring a one-to-one dynamical match between Arctic staircases and salt-finger route histories.}
  \label{fig:observable-state}
\end{figure}

\Cref{fig:observable-state-entropy,fig:observable-state-width} place the real and synthetic profiles in shared observable coordinates. The entropy-versus-correlation view separates scalar-gradient spread from scalar coupling. In the synthetic late-time table, entropy spans 0.684 to 0.806 and temperature-salinity correlation spans 0.804 to 0.941. The ITP profiles therefore overlap the synthetic scale in some scalar-gradient coordinates while showing higher median entropy and lower median temperature-salinity correlation. The salinity-IQR-versus-outer-fraction view asks whether a broadened gradient is also redistributed away from the central interface.

The ITP comparison is deliberately branch-aware. Arctic staircase profiles and finite-depth salt-fingering route histories are not interchangeable dynamical objects. The shared value is methodological: the same profile observables can be computed on real high-resolution profiles, and the resulting coordinate space exposes differences in spread, scalar coupling, and gradient localization. Those differences are useful precisely because they are not collapsed into a false route validation claim.

\FloatBarrier

\section{Routine Profiles and Hydrographic Sections}
\label{sec:routine_profiles_sections}

\subsection{Argo-Style Profile Sampling Stress-Tests Profile Observables}

The Argo-style analysis uses real autonomous-profile products to define local sampling records and applies those records to the synthetic profile problem. The screened pass contains 54 candidate profile records and yields 36 accepted local sampling records drawn from 22 unique profiles. The accepted local samples have a median interface pressure of 44.9 dbar, a median local sample size of 82.5 levels, and a median local pressure spacing of 1.98 dbar. Their median salinity-gradient width is 36.9 dbar and their median salinity-gradient entropy is 0.884.

When the accepted local sampling records are applied to the synthetic route histories, the median retained late synthetic route separation is 1.060 of native, with a 10--90 percent range of 0.964--1.221. Median relative salinity-gradient-IQR errors across route families range from 0.070 to 0.246, with a median across families of 0.137. These values give routine autonomous profiles a limited but useful role. They can stress-test observable quantities under real profile spacing and quality-control structure. They do not make the screened profile set a global Argo climatology, nor do they imply that Argo profiles recover exchange pathways directly.

\subsection{Sections Add Horizontal Information That Profiles Cannot Supply}

A section-like sampling reduction was applied to the synthetic fields to ask what horizontal information becomes available when station position is part of the observing format. The section quantities include an interface track, a dominant horizontal mode, and roughness relative to the native section. These quantities address a different part of the hidden route than profile bundles do. Profile bundles sample multiple local vertical structures. Sections can also sample the horizontal organization of the interface itself.

\begin{figure}[!htbp]
  \centering
  \begin{subfigure}[t]{0.49\linewidth}
    \centering
    \includegraphics[width=\linewidth]{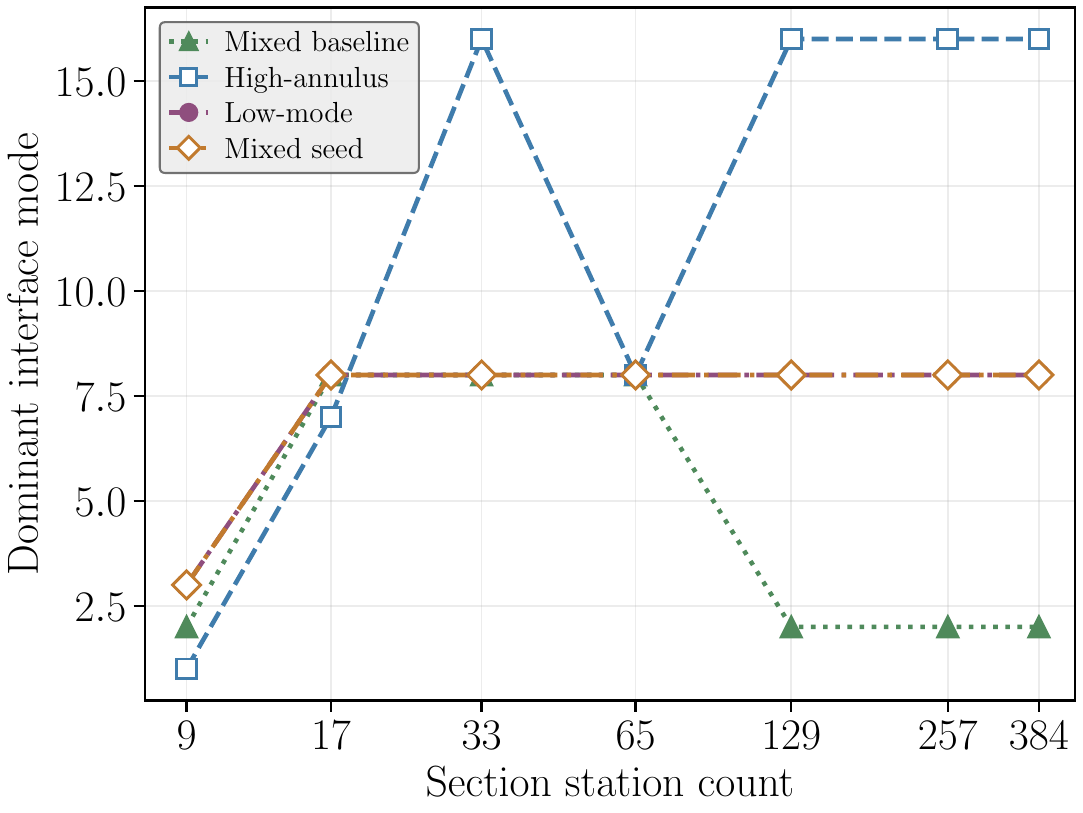}
    \caption{Dominant mode.}
    \label{fig:section-observability-mode}
  \end{subfigure}
  \hfill
  \begin{subfigure}[t]{0.49\linewidth}
    \centering
    \includegraphics[width=\linewidth]{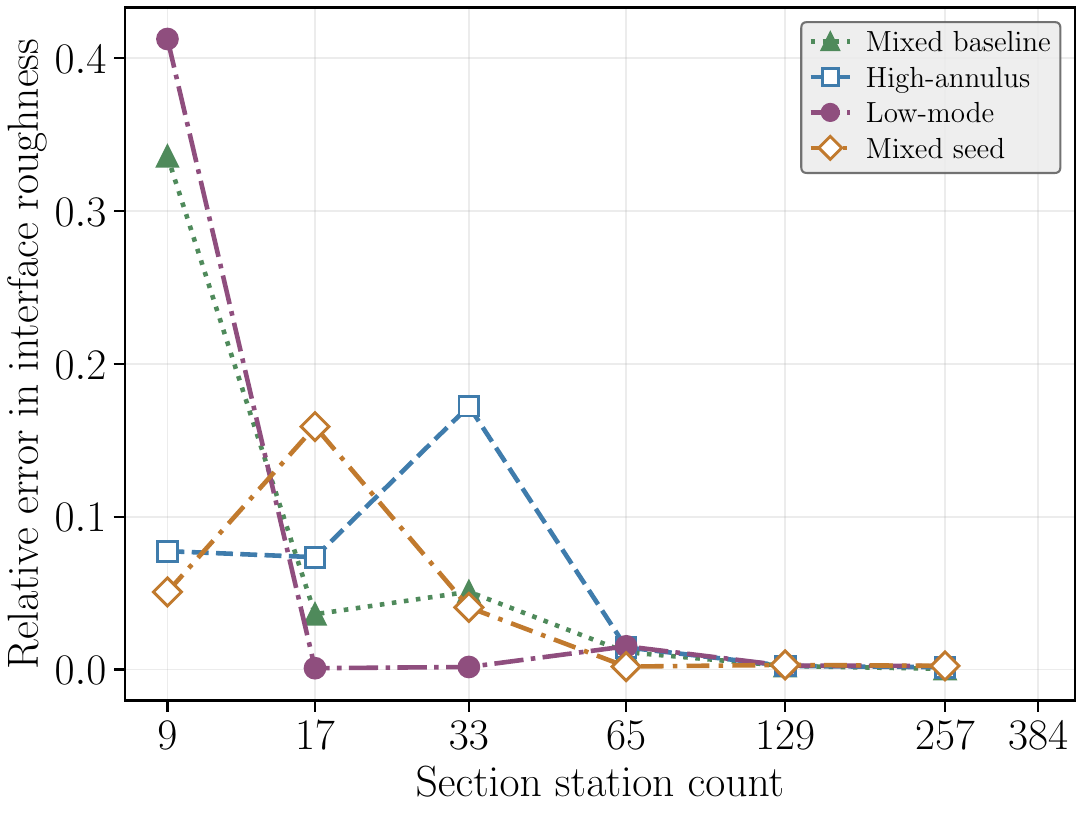}
    \caption{Roughness error.}
    \label{fig:section-observability-roughness}
  \end{subfigure}
  \caption{Section observability at the final comparison time. Station spacing controls whether horizontal modes can be recovered, while roughness can retain partial interface-amplitude information even when the dominant mode aliases.}
  \label{fig:section-observability}
\end{figure}

At the final comparison time, the native section-derived dominant modes across the four route histories are 2, 16, 8, and 8. These values are station-spacing limited. At 17 stations, the high-annulus mode aliases to 7 because the Nyquist mode is 8. Median section route separation is 0.858 of native at 17 stations and returns close to native by 65--257 stations. \Cref{fig:section-observability} presents the dominant-mode and roughness behavior.

The section result completes the observing hierarchy. A profile describes a local scalar interface. A profile bundle samples enough spatial variation to recover route-family structure. A section adds horizontal scale, but only if station spacing resolves that scale. Roughness preservation and dominant-mode recovery are not the same claim; a sparse section can retain some roughness information while misidentifying the dominant mode.

\begin{table}[!htbp]
  \centering
  \caption{Section-observability summary at the final comparison time. The table reports interface roughness, interquartile range, dominant mode, and spectral entropy for native and thinned section samplings.}
  \label{tab:section-observability}
  \small
  \begin{tabular}{llrrrr}
\toprule
Case & stations & roughness & IQR & dominant mode & entropy \\
\midrule
High-annulus & 17 & 13.727 & 22.038 & 7 & 0.707 \\
High-annulus & 33 & 17.371 & 20.900 & 16 & 0.665 \\
High-annulus & 65 & 15.032 & 18.914 & 8 & 0.605 \\
High-annulus & native & 14.817 & 19.477 & 16 & 0.380 \\
Low-mode & 17 & 17.333 & 25.845 & 8 & 0.642 \\
Low-mode & 33 & 17.287 & 26.437 & 8 & 0.679 \\
Low-mode & 65 & 17.581 & 23.343 & 8 & 0.610 \\
Low-mode & native & 17.317 & 24.122 & 8 & 0.435 \\
Mixed baseline & 17 & 15.651 & 21.784 & 8 & 0.809 \\
Mixed baseline & 33 & 17.068 & 27.292 & 8 & 0.730 \\
Mixed baseline & 65 & 16.425 & 24.653 & 8 & 0.647 \\
Mixed baseline & native & 16.240 & 24.794 & 2 & 0.431 \\
Mixed seed & 17 & 15.955 & 21.168 & 8 & 0.775 \\
Mixed seed & 33 & 14.326 & 19.693 & 8 & 0.851 \\
Mixed seed & 65 & 13.793 & 19.943 & 8 & 0.703 \\
Mixed seed & native & 13.766 & 20.063 & 8 & 0.487 \\
\bottomrule
\end{tabular}

\end{table}

\Cref{tab:section-observability} gives the corresponding roughness, interquartile range, dominant-mode, and entropy values. The compact high-annulus route is the clearest stress test for section spacing: its native dominant mode is 16 at the final comparison time, but a 17-station section cannot resolve that mode and aliases the estimate to 7. The low-mode and mixed histories are less demanding in this particular section metric because their dominant modes are lower, but they still show that station count and route geometry jointly determine how much horizontal information survives section thinning.

\FloatBarrier

\subsection{CCHDO/GO-SHIP Provides Real Section Bookkeeping}

The CCHDO/GO-SHIP I06S product provides a real section-scale context for station spacing, pressure spacing, and interface-track construction. The pilot section contains 109 profiles, of which 108 provide usable candidate-interface profiles under the minimal interface-tracking rule. The median station spacing is 55.4 km, the median strict-QC pressure spacing is 2.0 dbar, the median candidate-interface pressure is 88.6 dbar, and the median salinity-gradient width is 45.3 dbar. The median adjacent interface-pressure jump is 13.5 dbar, while the 90th percentile adjacent jump is 241.4 dbar.

This result establishes real-section readiness for interface-center tracking, gradient-width estimation, station-spacing assessment, and vertical-sampling assessment. The synthetic section results quantify how such section properties affect mode recovery and roughness fidelity for a route-known field. A field application would require branch and regime screening before making any specific double-diffusive route claim.

\section{Measurement Hierarchy and Implications}

The central result is that observability depends on the observing unit. An isolated profile can describe a local scalar transition but is not a robust identifier of the hidden finite-depth route. A profile bundle recovers route-family information more effectively because it samples spatial variation. A denser bundle is needed when the target is quantitative stability of multiple scalar-interface metrics. A section provides access to horizontal scale, but station spacing controls which modes can be interpreted.

This hierarchy explains why local profile metrics and route inference need to remain separate. Salinity-gradient width, central fraction, outer fraction, entropy, and temperature-salinity correlation describe the scalar transition that the instrument sampled. They do not identify whether that transition came from a compact route, a broad route, or a mixed route. Route information appears when multiple profiles or a section expose spatial organization.

The vertical-resolution results add a second design constraint. A coarse profile can preserve broad route-state separation under standardized metrics while distorting physical interface-width estimates. A survey can therefore be useful for broad state classification and still be inadequate for quantifying sharpness, outer-gradient reach, or metric-level interface fidelity. Observing design is clearest when it starts from the intended inference: local interface metric, broad route family, horizontal scale, or section roughness.

The real-data comparisons keep the synthetic results attached to practical measurement formats. ITP profiles show that high-resolution real profiles can be placed in the same scalar-observable space. Argo-style sampling records show how routine autonomous-profile spacing and quality-control structure perturb synthetic metrics. CCHDO/GO-SHIP sections show that real hydrographic products contain the station and vertical sampling needed for interface tracking. These connections are useful because they remain within their proper scope: they link measurement formats to observable quantities without claiming one-to-one field validation of idealized route histories.

The practical implication is that hidden double-diffusive exchange is best treated as an ensemble problem. A single cast is a local sample. A profile bundle can recover route-family information when the route has spatial structure. A section can identify horizontal organization if station spacing is adequate. Repeated sections, dense profile bundles, and combined profile-section strategies are therefore better suited to finite-depth exchange than isolated opportunistic profiles.

\section{Conclusions}

Sparse temperature-salinity measurements do not see finite-depth double-diffusive exchange in the same way that a resolved field does. The resolved field contains the hidden route history, while a profile measures a local scalar transition. In the controlled truth set analyzed here, isolated profiles remain ambiguous route indicators even late in the evolution.

Small profile bundles change the measurement problem. In the late-time bundle-enumeration framework, route-family recovery rises from a one-profile baseline of 0.694 to 0.917 for two-profile bundles, 0.961 for three-profile bundles, and 0.994 for four-profile bundles. This improvement indicates that the relevant observational unit for route-family inference is a local ensemble, not an individual cast.

Vertical sampling controls which scalar quantities can be used quantitatively. Coarsened profiles may retain broad route-state separation, but salinity-gradient width and related local interface measures can have large relative errors. Broad classification and local metric fidelity therefore need to be reported separately.

Sections add horizontal information that profiles cannot supply. The synthetic section analysis shows that station spacing can alias dominant horizontal modes even when some roughness information remains. Section design therefore matters both for mapping an interface and for deciding which horizontal scales can be interpreted.

The real profile and section products show how the same measurements can be summarized in practical observing formats. ITP profiles, Argo-style sampling records, and CCHDO/GO-SHIP sections provide high-resolution profile context, routine profile sampling realism, and section-scale geometry. Together with the synthetic truth reductions, they support a measurement hierarchy for finite-depth double-diffusive exchange: isolated profiles describe local state, profile bundles recover route-family information, and sections are needed for horizontal organization.

\appendix

\section{Vertical-Sampling Support}

\begin{figure}[!htbp]
  \centering
  \begin{subfigure}[t]{0.49\linewidth}
    \centering
    \includegraphics[width=\linewidth]{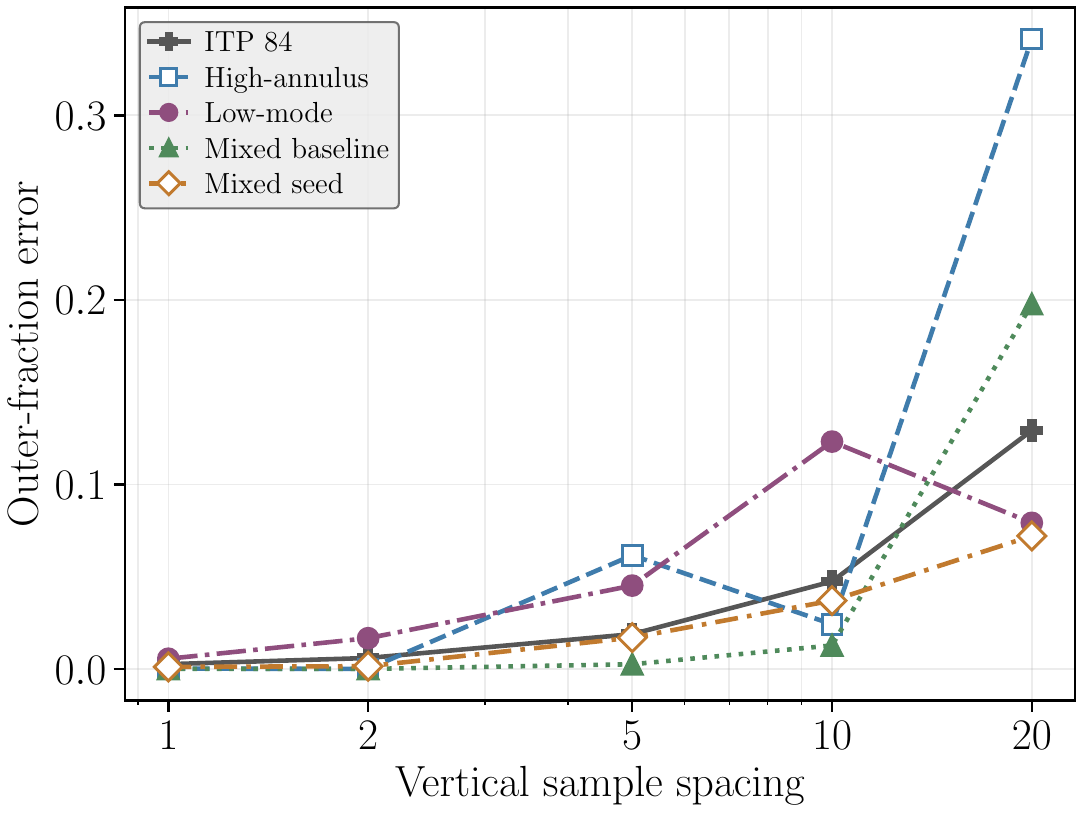}
    \caption{Outer-fraction error.}
    \label{fig:appendix-resolution-outer}
  \end{subfigure}
  \hfill
  \begin{subfigure}[t]{0.49\linewidth}
    \centering
    \includegraphics[width=\linewidth]{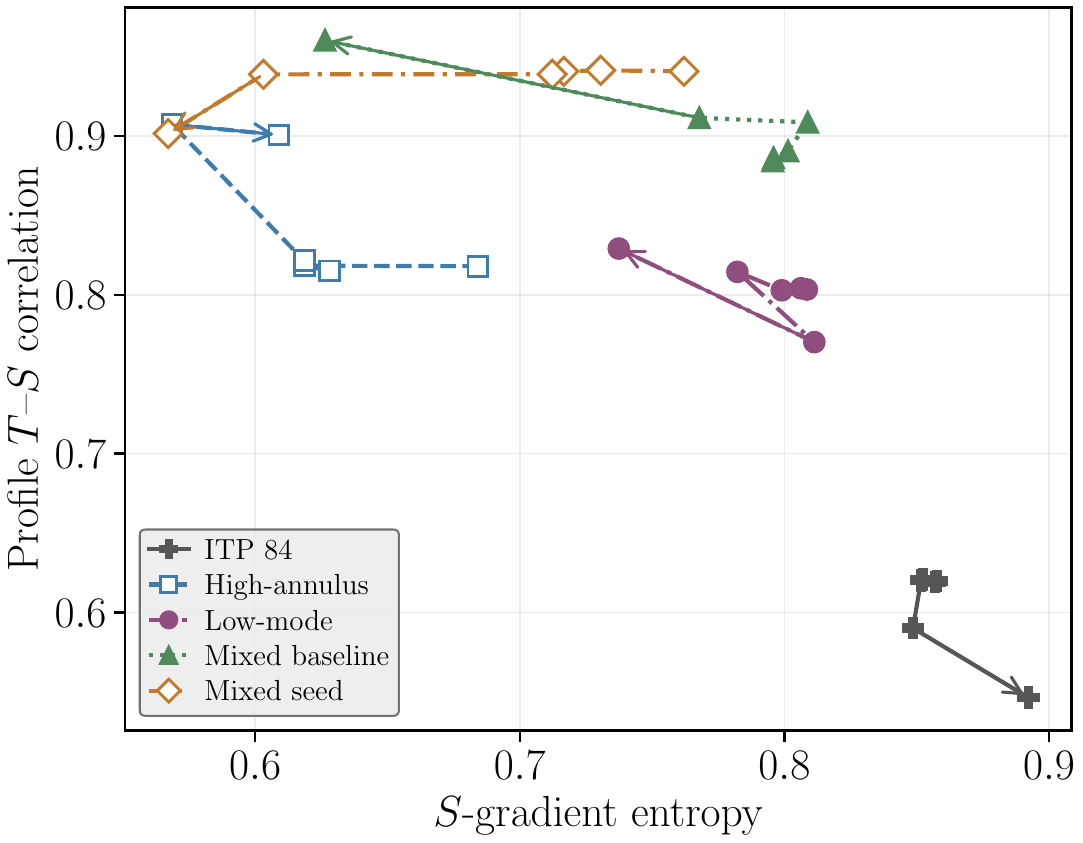}
    \caption{Observable-state trajectories.}
    \label{fig:appendix-resolution-trajectories}
  \end{subfigure}
  \caption{Supporting vertical-resolution quantities for profile observability.}
  \label{fig:appendix-resolution}
\end{figure}

The vertical-sampling appendix gathers supporting checks that separate broad state recovery from local metric fidelity. \Cref{fig:appendix-resolution} gives outer-fraction error and observable-state trajectories at the final comparison time, and \cref{tab:resolution-summary} records the numerical summary used by the resolution-degradation discussion.

\begin{table}[!htbp]
  \centering
  \caption{Resolution-degradation summary at the final comparison time.}
  \label{tab:resolution-summary}
  \small
  \begin{tabular}{llrrrr}
\toprule
Group & spacing & \(S\) IQR & outer frac. & \(S\) entropy & \(T\)--\(S\) corr. \\
\midrule
ITP 84 & native & 23.499 & 0.185 & 0.857 & 0.619 \\
ITP 84 & 5 & 24.280 & 0.203 & 0.852 & 0.620 \\
ITP 84 & 20 & 34.295 & 0.312 & 0.892 & 0.547 \\
High-annulus & native & 42.047 & 0.001 & 0.684 & 0.818 \\
High-annulus & 5 & 34.953 & 0.129 & 0.619 & 0.821 \\
High-annulus & 20 & 30.390 & 0.383 & 0.609 & 0.901 \\
Low-mode & native & 42.782 & 0.343 & 0.806 & 0.804 \\
Low-mode & 5 & 35.023 & 0.304 & 0.782 & 0.814 \\
Low-mode & 20 & 58.643 & 0.547 & 0.737 & 0.829 \\
Mixed baseline & native & 35.006 & 0.000 & 0.796 & 0.884 \\
Mixed baseline & 5 & 29.684 & 0.009 & 0.809 & 0.908 \\
Mixed baseline & 20 & 21.709 & 0.167 & 0.626 & 0.960 \\
Mixed seed & native & 24.893 & 0.004 & 0.762 & 0.941 \\
Mixed seed & 5 & 23.388 & 0.006 & 0.712 & 0.939 \\
Mixed seed & 20 & 20.907 & 0.325 & 0.567 & 0.901 \\
\bottomrule
\end{tabular}

\end{table}
\FloatBarrier

\section{Argo-Style Profile Support}

\begin{figure}[!htbp]
  \centering
  \begin{subfigure}[t]{0.49\linewidth}
    \centering
    \includegraphics[width=\linewidth]{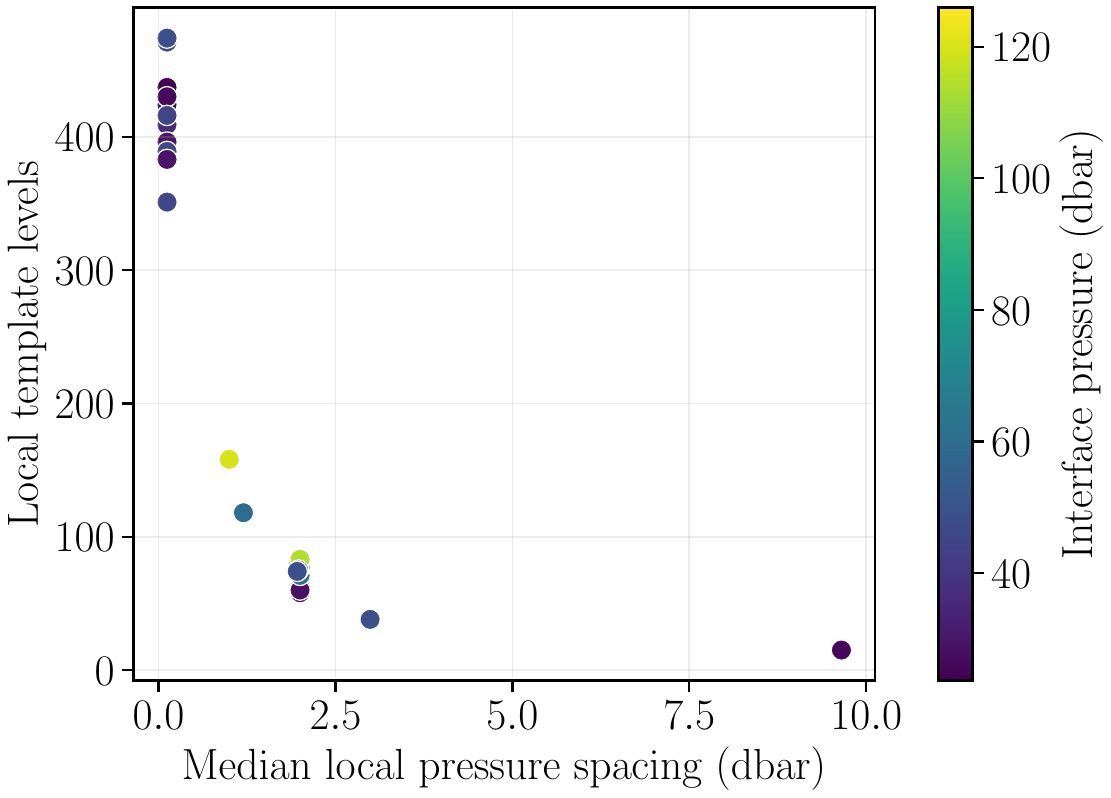}
    \caption{Accepted local-sample distribution.}
    \label{fig:appendix-argo-distribution}
  \end{subfigure}
  \hfill
  \begin{subfigure}[t]{0.49\linewidth}
    \centering
    \includegraphics[width=\linewidth]{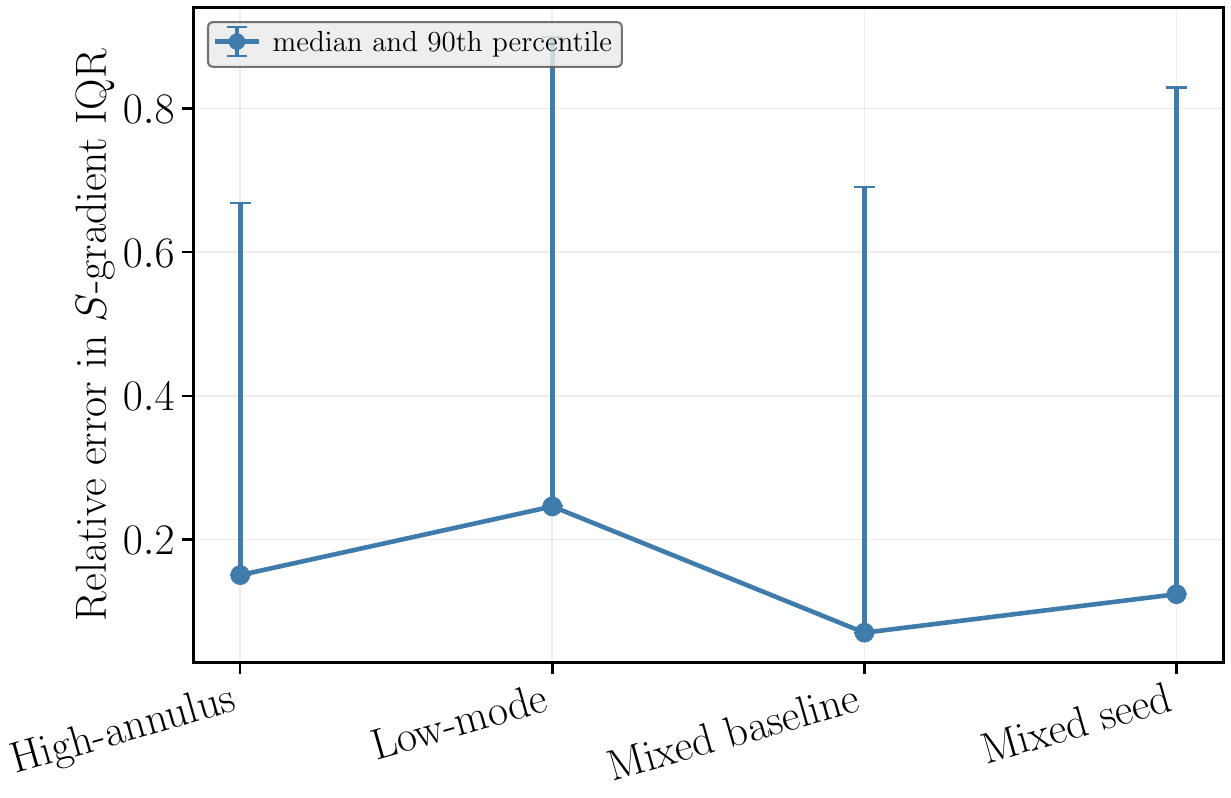}
    \caption{Salinity-gradient-IQR error.}
    \label{fig:appendix-argo-iqr}
  \end{subfigure}
  \caption{Argo-style support material for autonomous-profile sampling.}
  \label{fig:appendix-argo}
\end{figure}

The Argo-style appendix records the accepted local-sample distribution and the synthetic metric perturbations obtained after applying autonomous-profile sampling records. \Cref{fig:appendix-argo} provides the supporting visual check; the key population and perturbation values are reported in \cref{sec:routine_profiles_sections}.

\FloatBarrier

\section{CCHDO/GO-SHIP Section Support}

\begin{figure}[!htbp]
  \centering
  \begin{subfigure}[t]{0.49\linewidth}
    \centering
    \includegraphics[width=\linewidth]{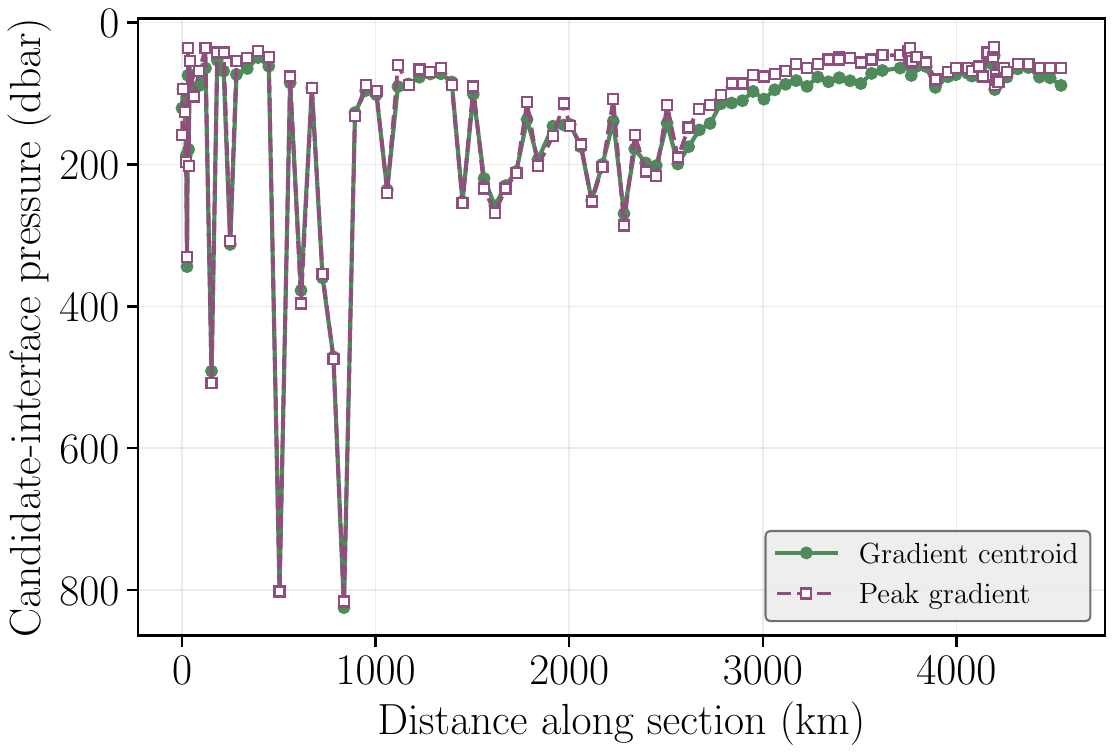}
    \caption{Candidate-interface track.}
    \label{fig:appendix-cchdo-track}
  \end{subfigure}
  \hfill
  \begin{subfigure}[t]{0.49\linewidth}
    \centering
    \includegraphics[width=\linewidth]{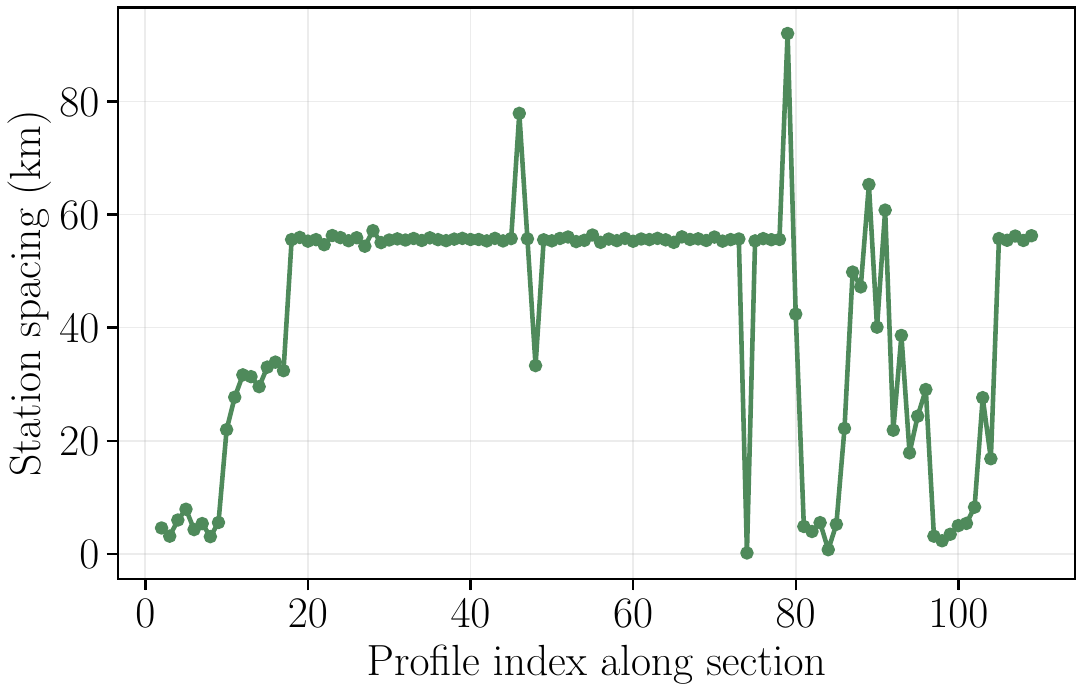}
    \caption{Station spacing.}
    \label{fig:appendix-cchdo-spacing}
  \end{subfigure}
  \caption{CCHDO/GO-SHIP I06S section support material.}
  \label{fig:appendix-cchdo}
\end{figure}

The CCHDO/GO-SHIP appendix records the section geometry that connects hydrographic-station spacing to the synthetic section sampling. \Cref{fig:appendix-cchdo} gives the real-section interface track and station spacing; the compact numerical summaries are reported in \cref{sec:routine_profiles_sections}.

\FloatBarrier

\section*{Statements and Declarations}

\noindent\textbf{Funding.}
This research received no specific grant from any funding agency in the public, commercial, or not-for-profit sectors.

\noindent\textbf{Competing interests.}
The author declares no competing interests.

\noindent\textbf{Author contributions.}
Sriram P. Kalathoor performed the conceptualization, methodology, software development, data curation, analysis, visualization, and writing.

\noindent\textbf{Data availability.}
The finite-depth simulation products used to construct the synthetic-observation reductions are derived from the archived data record associated with the companion studies \citep{kalathoor2026swfdata}. Public Ice-Tethered Profiler, Argo, and CCHDO/GO-SHIP products used for observing-format context are available from their respective program archives. The processed tabular products produced for this study will be archived in a public repository before publication; interim copies are available from the corresponding author on reasonable request.

\noindent\textbf{Use of generative AI and AI-assisted technologies.}
During preparation of this work, AI-assisted tools were used for language editing, code review, and manuscript organization. The author reviewed and edited the resulting material and takes full responsibility for the content of this article.

\ifspringerclass\else
  \bibliographystyle{plainnat}
\fi
\bibliography{references}

\end{document}